\documentclass[journal]{IEEEtran}

\usepackage[mathscr]{eucal}
\usepackage[cmex10]{amsmath}
\usepackage{epsfig,epsf,psfrag}
\usepackage{amssymb,amsmath,amsthm,amsfonts,latexsym}
\usepackage{amsmath,graphicx,bm,xcolor,url}
\usepackage[caption=false]{subfig} 
\usepackage{fixltx2e}
\usepackage{array}
\usepackage{verbatim}
\usepackage{bm}
\usepackage{algorithmic, cite}
\usepackage{algorithm}
\usepackage{verbatim}
\usepackage{textcomp}
\usepackage{mathrsfs}
\usepackage{epstopdf}

\catcode`~=11 \def\UrlSpecials{\do\~{\kern -.15em\lower .7ex\hbox{~}\kern .04em}} \catcode`~=13 

\allowdisplaybreaks[3]

\newcommand{\ba}{\mathbf{a}}

\newcommand{\bB}{\mathbf{B}}

\newcommand{\be}{\mathbf{e}}

\newcommand{\bI}{\mathbf{I}}

\newcommand{\bp}{\mathbf{p}}

\newcommand{\bQ}{\mathbf{Q}}

\newcommand{\bs}{\mathbf{s}}

\newcommand{\bT}{\mathbf{T}}

\newcommand{\bU}{\mathbf{U}}

\newcommand{\bz}{\mathbf{z}}

\DeclareMathAlphabet{\mathbsf}{OT1}{cmss}{bx}{n}
\DeclareMathAlphabet{\mathssf}{OT1}{cmss}{m}{sl}

\DeclareSymbolFont{bsfletters}{OT1}{cmss}{bx}{n}  
\DeclareSymbolFont{ssfletters}{OT1}{cmss}{m}{n}
\DeclareMathSymbol{\bsfGamma}{0}{bsfletters}{'000}
\DeclareMathSymbol{\ssfGamma}{0}{ssfletters}{'000}
\DeclareMathSymbol{\bsfDelta}{0}{bsfletters}{'001}
\DeclareMathSymbol{\ssfDelta}{0}{ssfletters}{'001}
\DeclareMathSymbol{\bsfTheta}{0}{bsfletters}{'002}
\DeclareMathSymbol{\ssfTheta}{0}{ssfletters}{'002}
\DeclareMathSymbol{\bsfLambda}{0}{bsfletters}{'003}
\DeclareMathSymbol{\ssfLambda}{0}{ssfletters}{'003}
\DeclareMathSymbol{\bsfXi}{0}{bsfletters}{'004}
\DeclareMathSymbol{\ssfXi}{0}{ssfletters}{'004}
\DeclareMathSymbol{\bsfPi}{0}{bsfletters}{'005}
\DeclareMathSymbol{\ssfPi}{0}{ssfletters}{'005}
\DeclareMathSymbol{\bsfSigma}{0}{bsfletters}{'006}
\DeclareMathSymbol{\ssfSigma}{0}{ssfletters}{'006}
\DeclareMathSymbol{\bsfUpsilon}{0}{bsfletters}{'007}
\DeclareMathSymbol{\ssfUpsilon}{0}{ssfletters}{'007}
\DeclareMathSymbol{\bsfPhi}{0}{bsfletters}{'010}
\DeclareMathSymbol{\ssfPhi}{0}{ssfletters}{'010}
\DeclareMathSymbol{\bsfPsi}{0}{bsfletters}{'011}
\DeclareMathSymbol{\ssfPsi}{0}{ssfletters}{'011}
\DeclareMathSymbol{\bsfOmega}{0}{bsfletters}{'012}
\DeclareMathSymbol{\ssfOmega}{0}{ssfletters}{'012}

\newcommand{\qednew}{\nobreak \ifvmode \relax \else
      \ifdim\lastskip<1.5em \hskip-\lastskip
      \hskip1.5em plus0em minus0.5em \fi \nobreak
      \vrule height0.75em width0.5em depth0.25em\fi}

\begin{document}

\title{Sparsity-Based Error Detection in \\ DC Power Flow State Estimation}
\author{M.H.~Amini$^{1,2,3,4}$, \textit{Graduate Student Member, IEEE}, Mostafa Rahmani, Kianoosh~G.~Boroojeni$^6$, \\ George Atia$^{5}$, \textit{Member, IEEE}, and S.S.~Iyengar$^6$, \textit{Fellow, IEEE}, and O. Karabasoglu$^{2,3,4}$, \textit{Member, IEEE}\\ \small $^1$ Department of Electrical and Computer Engineering, Carnegie Mellon University, Pittsburgh, PA, USA\\$^2$ SYSU-CMU Joint Institute of Engineering, Pittsburgh, PA, USA\\$^3$ SYSU-CMU Shunde International, Joint Research Institute, Guangdong, China\\ $^4$ School of Electronics and Information Technology, SYSU, Guangzhou, China\\ $^5$Department of Electrical Engineering and Computer Science, University of Central Florida, Orlando,
 FL, USA \\ $^6$ School of Computing and Information Sciences, Florida International
 University, Miami, FL, USA  
} 
\maketitle
\begin{abstract}

This paper presents a new approach for identifying the measurement error in the DC power flow state estimation problem. The proposed algorithm exploits the singularity of the impedance matrix and the sparsity of the error vector by posing the DC power flow problem as a sparse vector recovery problem that leverages the structure of the power system and uses $l_1$-norm minimization for state estimation. This approach can provably compute the measurement errors exactly, and its performance is robust to the arbitrary magnitudes of the measurement errors. Hence, the proposed approach can detect the noisy elements if the measurements are contaminated with additive white Gaussian noise plus sparse noise with large magnitude. 
The effectiveness of the proposed sparsity-based decomposition-DC power flow approach is demonstrated on the IEEE 118-bus and 300-bus test systems.
\end{abstract}


\section{Introduction}
Emerging technologies, including advanced metering infrastructure (AMI), phasor measurement units (PMUs), and distributed renewable energy resources brought to the fore the concept of the smart grid (SG) as a novel means for advancing towards more secure, reliable, sustainable, and environmentally-friendly power systems \cite{giana,amini13}. The SG can be defined as electric power networks enhanced by information communications technology (ICT)\cite{marija1} as a result of integrating advanced control
and communication technologies with the traditional power grid \cite{deepa}.

The integration of distributed renewable energy resources is also one of the challenges of future power systems \cite{mahdi1}. Distributed generation placement has been addressed in \cite{mohsennn}.  Distributed optimization is utilized to improve the computational aspects of power systems, e.g. two decomposition techniques are evaluated in \cite{amini15}. 
In \cite{vpoor}, a branch and bound algorithm was used to determined globally optimal locations of sensors. The performance of the SG is assessed based on several important studies, including stability analysis, power flow calculation, and reliability analysis \cite{moslehisgreli}. A conceptual SG model based on seven specific domains was proposed by the National Institute of Standards and Technology \cite{NIST}.

Optimization methods are crucial to future power systems. This includes particle swarm optimization, which is widely used for power systems optimization \cite{arman2}.  According to \cite{rahimisg}, recent equipment at the demand and supply sides will improve the reliability of future power systems. In this context, demand response programs play a pivotal role in helping the utility companies and suppliers reduce the system cost \cite{aminisgj}.
A semidefinite relaxation-based nonlinear state estimation approach was introduced in \cite{zhu} to achieve near-optimal performance of power networks. The proposed approach has the advantage of reducing the computational complexity owing to distributed implementation. In a related context, power flow studies can take advantage of signal processing measures to facilitate smart grid operations \cite{gianareview}. For instance, in \cite{amini2014} the minimum-variance unbiased estimator (MVUE) was determined for DC power flow estimation. Consequently, there is a pressing need to improve the accuracy of power flow solutions and reduce the computational burden.
Wide utilization of cloud computing applications for big data analytics is necessary to deal with the heterogeneous structure of the SG \cite{cloud1,cloud2}. Additionally, big data analytics have been used to ameliorate both power system operation and protection \cite{bigdata1}.

The DC power flow (DCPF) model hinges on simplifying assumptions -- to be explained later -- and is used for power system monitoring and optimization at the transmission level. 
Compared to AC power flow methods, DCPF has some advantages, namely, providing a unique solution, reliability of the outputs, straightforward and non-iterative implementation, adaptability in power system optimization, and maintaining justifiable accuracy \cite{dcstot,dchadi}. Additionally, due to the convexity of the DCPF problem, the optimality of its distributed solution can be established \cite{moha2}.

In this paper, we build on the DCPF study by introducing a noise detection approach to improve the accuracy of power flow estimation. The proposed method, termed SD-DCPF (Sparsity-based Decomposition-DCPF), is based on an $l_1$-norm minimization problem. It is assumed that the measurement noise is a summation of natural noise and sparse noise with arbitrary magnitude. The large magnitude measurement errors may be intentionally injected into the system. We exploit the sparsity of the dominant elements of the measurement noise vector and the singularity of the impedance matrix in order to decompose the measurement vector into true measurements and an additive error vector. We show that exact decomposition is guaranteed if the impedance matrix satisfies certain conditions. Based on the extracted noise vector, accurate power flow solutions are determined.

 The rest of this paper is organized as follows. In Section II, the problem is defined and the DCPF formulation is introduced. In Section III, SD-DCPF is discussed. Section IV presents a numerical study using noisy measurement data for IEEE 118-bus and 300-bus test cases. Section V concludes the paper.

\section{Background and Problem Formulation}
In this section, we introduce the sparsity-based decomposition algorithm and discuss the simplifying assumptions underlying DCPF.

\subsection{Sparse Vector Recovery}
Many signals and data of interest exhibit low-dimensional structures that are often ignored by classic data analysis techniques. For instance, in some applications we may be interested in recovering a vector that is inherently sparse (or has a sparse representation in some basis) or a matrix that is naturally low rank.
A major line of research has focused on developing new approaches to leverage these low-dimensional structures \cite{lamport3 , mpaper,mpaper1n}. For instance, it was shown that sparse vectors can be recovered from a small number of non-adaptive measurements. Hence, the sparsity of a signal can be exploited to recover it from significantly less measurements than required by the Nyquist sampling rate \cite{lamport12,lamport3}.
Also, randomized robust subspace recovery is utilized with high-dimensional data matrices in \cite{mpaper2}.

A powerful result was derived in \cite{lamport3} showing that a sparse vector can be recovered from a small set of random orthogonal projections. Suppose that $\bs \in \mathbb{R}^{N}$ is a sparse vector and the number of its non-zero elements is equal to $\| \bs \|_0$. Define $\bT \in \mathbb{R}^{N \times N}$ as an orthonormal basis, i.e.,
\begin{eqnarray}
\bT^T \bT = \bI,
\end{eqnarray}
where $\bI$ is the identity matrix. Now suppose that $\bU \in \mathbb{R}^{N \times m}$ is formed from a set of randomly chosen columns of $\bT$. In addition, define $\mu$ as the minimum value which satisfies
\begin{eqnarray}
\underset{i}{\max} \|\bU^T\be_i\|_2 \leq \frac{\mu m}{N}.
\label{inch}
\end{eqnarray}
If the parameter $\mu$ is small, then the columns subspace of $\bU$ is not aligned with the standard basis. The orthogonal matrix $\bU$ is used to measure the sparse vector, thus it is essential that $\bU$ is not a sparse matrix.
In \cite{lamport3}, it was shown that if
\begin{eqnarray}
m \ge c \| \bs \|_0 \mu \log \frac{N}{\delta},
\label{spc}
\end{eqnarray}
then the optimal point of
\begin{eqnarray}
\begin{aligned}
& \underset{\hat{\bz}}{\min}
& & \| \hat{\bz}\|_1  \\
& \text{subject to}
& & \bU^T \hat{\bz} = \bU^T \bs .
\end{aligned}
\end{eqnarray}
is equal to $\bs$ with probability as least $(1 - \delta)$, where $c$ is a constant number. Hence, an $N$-dimensional sparse vector can be recovered from a small number of non-adaptive random linear measurements, and the sufficient number of such measurements scales linearly with the number of non-zero elements of the sparse vector (or the number of dominant elements if it is not exactly sparse \cite{lamport13}).

\subsection {DC Power Flow Problem}
Assume that the transmission line between the $i^\text{th}$ and $j^\text{th}$ buses is represented by its impedance $\mathcal{Z}_{ij}=R_{ij}+jX_{ij}$, where $R_{ij}$ and $X_{ij}$ represent the resistance and reactance of line $ij$, respectively. We can calculate the admittance of line $ij$ as $\mathcal{Y}_{ij}=1/\mathcal{Z}_{ij}=g_{ij}+jb_{ij}$. Then, the active power flow of this line is calculated using the following equation,
\[
P_{ij}=V_{i}^2g_{ij}-V_{i}V_{j}g_{ij}\cos(\delta_i-\delta_j)+V_{i}V_{j}b_{ij}\sin(\delta_i-\delta_j),
\]
where $\delta_i$ and $V_i$ represent the voltage angle and voltage magnitude at the $i^\text{th}$ bus, respectively. The DCPF model is based on the following three assumptions.

\begin{enumerate}
\item In the high voltage transmission line, the resistive part of the impedance can be neglected because of the large value of $X_{ij}/R_{ij}$ \cite{gianareview}. Therefore, the network is fairly inductive, i.e. $R_{ij}=0$.
\item The voltage angle of two connected buses is very small ($\delta_i-\delta_j \approx 0$). Therefore, we approximate the trigonometric functions, i.e. $\cos (\delta_i-\delta_j) \approx 1$ and $\sin (\delta_i-\delta_j) \approx 0$.
\item According to the normal operation conditions of the power system, the voltage magnitude is replaced by one p.u.
\end{enumerate}

Based on the aforementioned assumptions, the active power flow of line $ij$ is calculated using $P_{ij}=\frac{\delta_i-\delta_j}{X_{ij}}$. Let us define $\textbf{p}$ as the vector of active power values and $\boldsymbol {\delta}$ as the vector of voltage phases. We define $\textbf{p}= \textbf{B}  \boldsymbol{\delta}$, where the entries of $\textbf{B}$ are obtained by
\[\textbf{B}_{ij}=
  \begin{cases}
    \sum\limits_{k \in {\cal T}\setminus\{i\}}
                {X_{ik}}^{-1} & ; ~i=j,~~ \forall i \in {\cal T} \\
   -X_{ij}^{-1} & {; ~i\neq j,~ \text{line}~ ij~ \text{exists}}, i,j \in {\cal T}  \\
   0 & ;~ \text{otherwise}
  \end{cases}
  \]
where ${\cal T}$ denotes the set of all buses. In order to model the noise vector, we update the DCPF model to
\begin{eqnarray}
\bp =\bB \boldsymbol{\delta} + \mathbf{\epsilon},
          \label{eq:main}
\end{eqnarray}
where $\boldsymbol{\epsilon}$ denotes the noise vector. We are considering the regime of small number of noisy measurements, i.e., measurements are highly-accurate at the transmission system. We exploit this geometrical structure to ameliorate the quality of the Sparse Vector Recovery algorithm.

\section{Proposed SD-DCPF}
Due to the existence of a slack bus, we have at least a row that is linearly dependent on the other rows of the $\bB$ matrix; hence the matrix $\bB$ is not a full rank matrix in most configurations. Suppose that the rank of the matrix $\bB$ is equal to $r_B$. Then, (\ref{eq:main}) can be rewritten as
\begin{eqnarray}
\bp = \bQ \mathbf{a} + \mathbf{\epsilon},
\end{eqnarray}
where $\bQ \in \mathbb{R}^{\aleph \times r_B}$ is an orthonormal basis for the columns subspace of $\bB$. The conventional method for estimating the coefficient vector $\ba$ is the least-squares method, which solves the following optimization problem
\begin{eqnarray}
\underset{\hat{\ba}}{\min} \, \, \| \bp - \bQ \hat{\ba} \|_2.
\label{lsq}
\end{eqnarray}
The least-squares approach to (\ref{lsq}) projects the vector $\bp$ onto the columns subspace of $\bQ$. Therefore, the quality of the least-square estimator is function of the projection of the noise vector $\mathbf{\epsilon}$ on the columns subspace of $\bQ$. As a result, the least squares approach does not yield accurate estimation if the noise vector has a large projection on the columns subspace of $\bQ$.

In this paper, we consider scenarios in which the error vector $\mathbf{\epsilon}$ is relatively sparse. In contrast to $l_2$-minimization algorithms, $l_1$-minimization algorithms are robust to additive sparse errors with arbitrarily large magnitudes \cite{lamport13 , lamport17}. In particular, if $\mathbf{\epsilon}$ is sufficiently sparse and $\bQ$ satisfies the incoherence conditions \cite{lamport17 , mpaper}, then the optimal point of
\begin{eqnarray}
\underset{\hat{\ba}}{\min} \, \, \| \bp - \bQ \hat{\ba} \|_1
\label{l1}
\end{eqnarray}
is equal to $\ba$, and the optimal point does not depend on the magnitude of $\mathbf{\epsilon}$.

In this paper, we assume that $\bB$ is not a full rank matrix, which according to our investigations prevails many network structures. Thus, $r_B < \aleph $. Let us define $\bQ^{\perp} \in \mathbb{R}^{\aleph \times (\aleph  - r_B)}$ as the matrix whose columns subspace is the complement of the columns subspace of $\bQ$. It was shown in \cite{lamport17 , mpaper} that (\ref{l1}) is equivalent to
\begin{eqnarray}
\begin{aligned}
& \underset{\hat{\mathbb{\epsilon}}}{\min}
& & \| \hat{\mathbb{\epsilon}}\|_1  \\
& \text{subject to}
& & (\bQ^{\perp})^T \hat{\mathbb{\epsilon}} = (\bQ^{\perp})^T \bp,
\end{aligned}
\label{sr}
\end{eqnarray}
to say that, if $\ba_o$ is the optimal point of (\ref{l1}) and $\hat{\mathbb{\epsilon}}_o$ is the optimal point of (\ref{sr}), then $\bp - \bQ\ba_o = \hat{\mathbb{\epsilon}}_o$. Therefore, according to (\ref{spc}), if
\begin{eqnarray}
(\aleph  - r_B) \ge c \| \mathbb{\epsilon}\|_0 \mu_B \log \frac{N}{\delta}
\end{eqnarray}
then the optimal point of (\ref{sr}) is equal to $\ba$ with probability at least $(1-\delta)$, where $\| \mathbb{\epsilon}\|_0$ is the number of non-zero elements of $\mathbb{\epsilon}$, and similar to (\ref{inch}), $\mu_B$ is defined as follows,
\begin{eqnarray}
\underset{i}{\max} \|(\bQ^{\perp})^T\be_i\|_2 \leq \frac{\mu_B (\aleph  - r_B)}{\aleph}.
\end{eqnarray}

Therefore, if the rank of $\bB$ is sufficiently small (i.e., $(\aleph  - r_B)$ is large enough), the $l_1$-minimization algorithm (\ref{l1}) can yield exact estimation.

Henceforth, we utilize the aforementioned approach to detect the noise vector $\epsilon$ in (\ref{eq:main}). Our approach is proposed for power flow calculation in transmission networks with measurements that are accurate for the most part, i.e., the noise vector is sparse. Consequently, SD-DCPF refers to solving the DCPF problem as a linear problem with noisy measurement inputs utilizing sparsity-based vector decomposition. It also finds the noisy measured data to increase the accuracy of the DCPF solution.

 \section{Case Study and Discussion}
In this section, the performance of the proposed algorithm is analyzed. It is shown that the proposed $\ell_1$-norm minimization algorithm yields robustness against the large values of the sparse errors. We apply the algorithm to IEEE 118-bus and 300-bus test case systems \cite{tests}. First, it is observed that the performance of the algorithm is function of the number of dominant elements (or non-zero elements). Then, in contrast to the least-squares approach, the estimation accuracy of the proposed algorithm is shown to be independent of the values of the dominant elements.
\subsection{Singularity of the Matrix B}
In this section, we study the singularity of the $\bB$ matrix. Fig. 1 shows the singular values of the $\bB$ matrix for a 300-bus system. It is evident that a remarkable portion of the singular values are zero or negligible. Thus, the $\bB$ matrix is a singular matrix.
       \begin{figure}[h!]
   	\centering
       \includegraphics[width=0.5\textwidth]{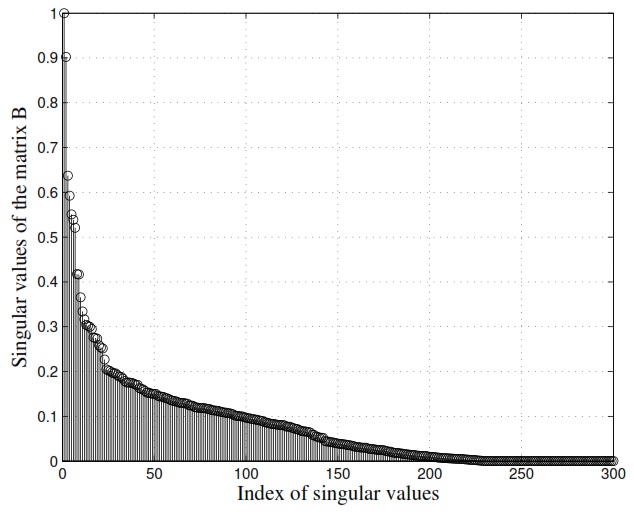}
       \vspace{-.1in}
       \caption{Singular values of the $\bB$ matrix for the IEEE 300-bus system.}

   \end{figure}
    \subsection{IEEE 118-Bus Test Case}
Here, we study the performance of the proposed algorithm under different sparsity levels. Suppose that the additive error vector is a sparse vector. Each element of the error vector is non-zero with probability equal to $\alpha$. Thus, we control the sparsity level by adjusting the parameter $\alpha$.  Figures 2-4 represent the estimated error vector for three different values of $\alpha$. One can see that the performance of the proposed algorithm improves when the number of non-zero elements of the error vector is decreased. This agrees with the analysis, which shows that the proposed estimator is only sensitive to the number of dominant elements (or non-zero elements) of the error vector.

   \begin{figure}[h!]
   	\centering
       \includegraphics[width=0.48\textwidth]{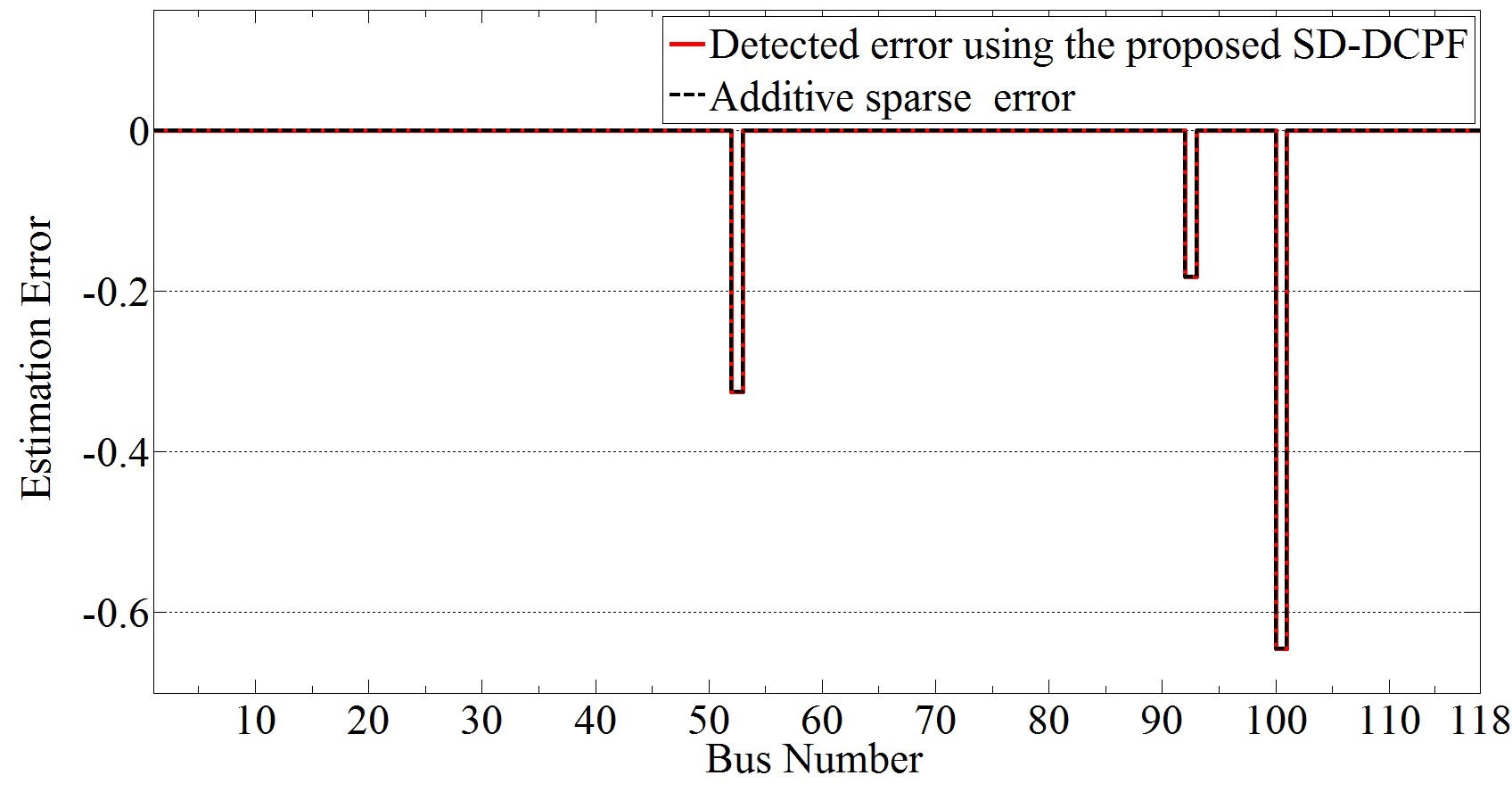}
       \vspace{-.1in}
       \caption{Detected estimation error, \textit{Scenario {I}}, $\alpha=0.03$.}
       \label{fig1}
   \end{figure}

   \begin{figure}[h!]
   	\centering
       \includegraphics[width=0.48\textwidth]{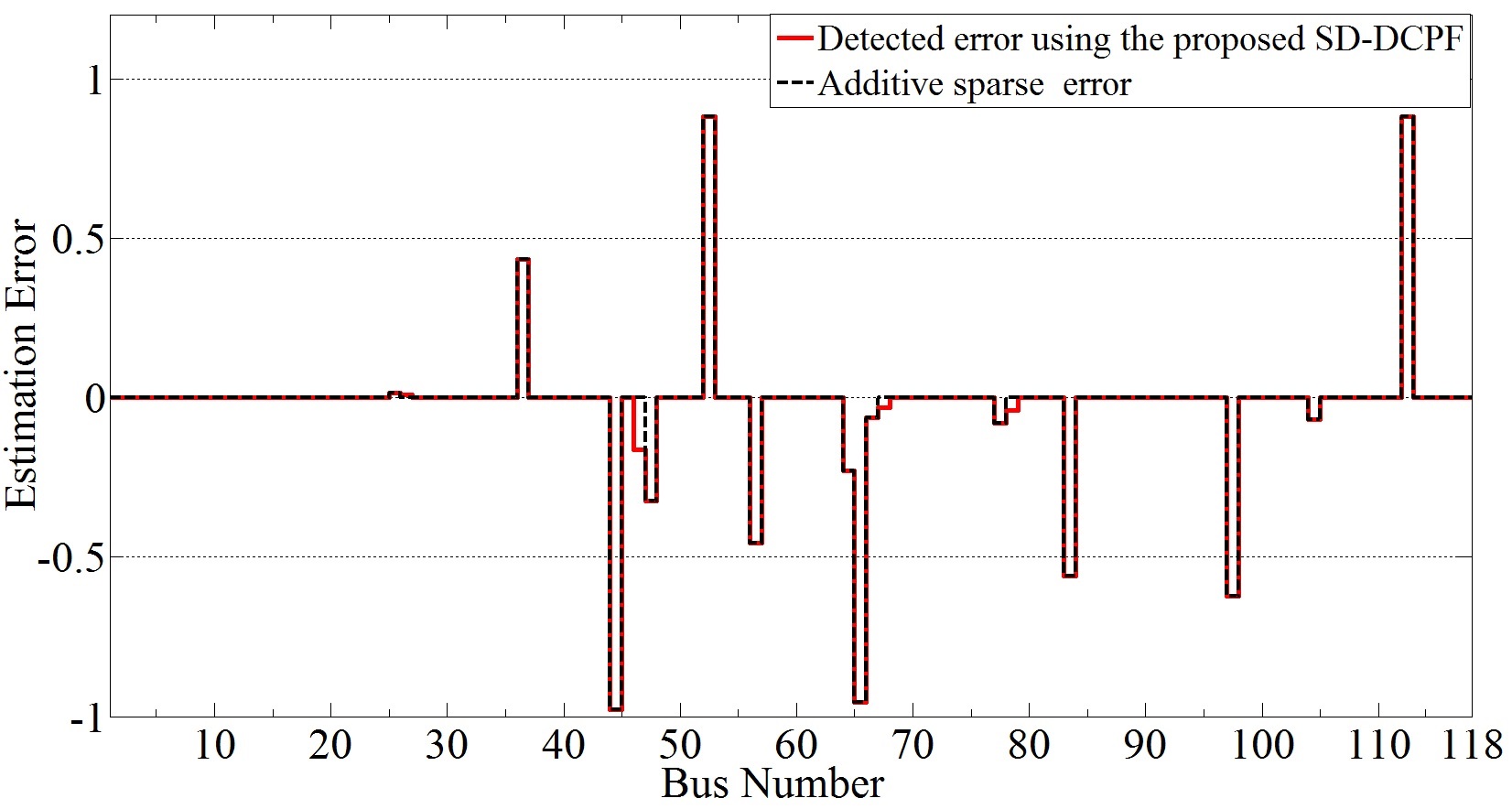}
       \vspace{-.1in}
       \caption{Detected estimation error, \textit{Scenario {II}}, $\alpha=0.08$.}
       \label{fig2}
   \end{figure}

      \begin{figure}[h!]
      	\centering
          \includegraphics[width=0.48\textwidth]{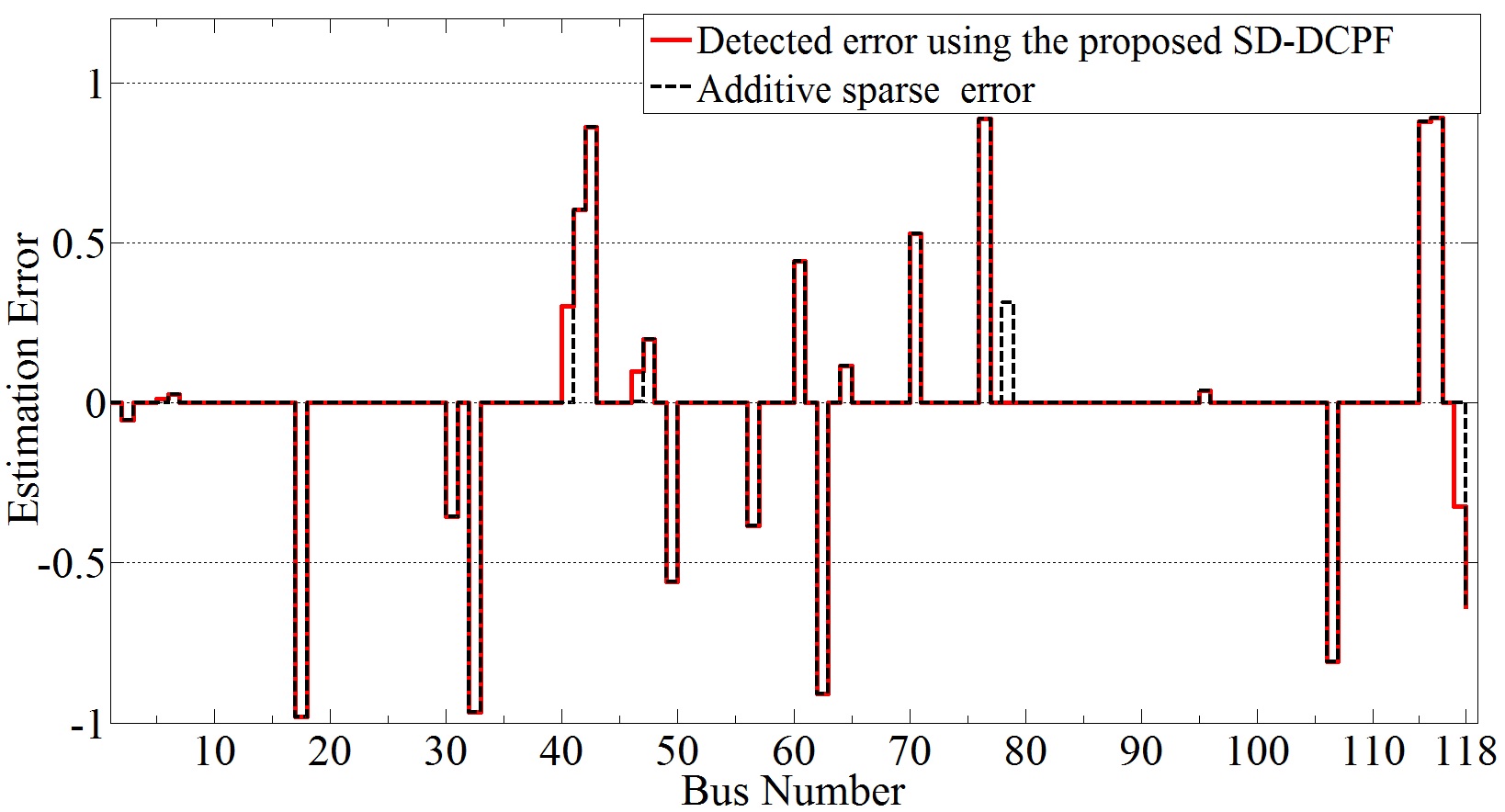}
          \vspace{-.1in}
          \caption{Detected estimation error, \textit{Scenario {III}}, $\alpha=0.15$.}
          \label{fig3}
      \end{figure}

   \subsection{IEEE 300-Bus Test Case}
 Now we apply the proposed algorithm to the 300-bus system. We show that the proposed approach is not sensitive to the magnitudes of the elements of the sparse error vector. Three scenarios are defined. Scenario IV is defined not only to evaluate the scalability of the proposed approach, but also to evaluate the accuracy of our approach on different test networks. In these scenarios, we use $\alpha$ equal to $8\%$. Scenario V is used to compare the performance of our approach and the least-square estimator (LSE). We also define scenario VI to evaluate the effect of noise magnitude on performance.

   \begin{figure}[h!]
   	\centering
       \includegraphics[width=0.48\textwidth]{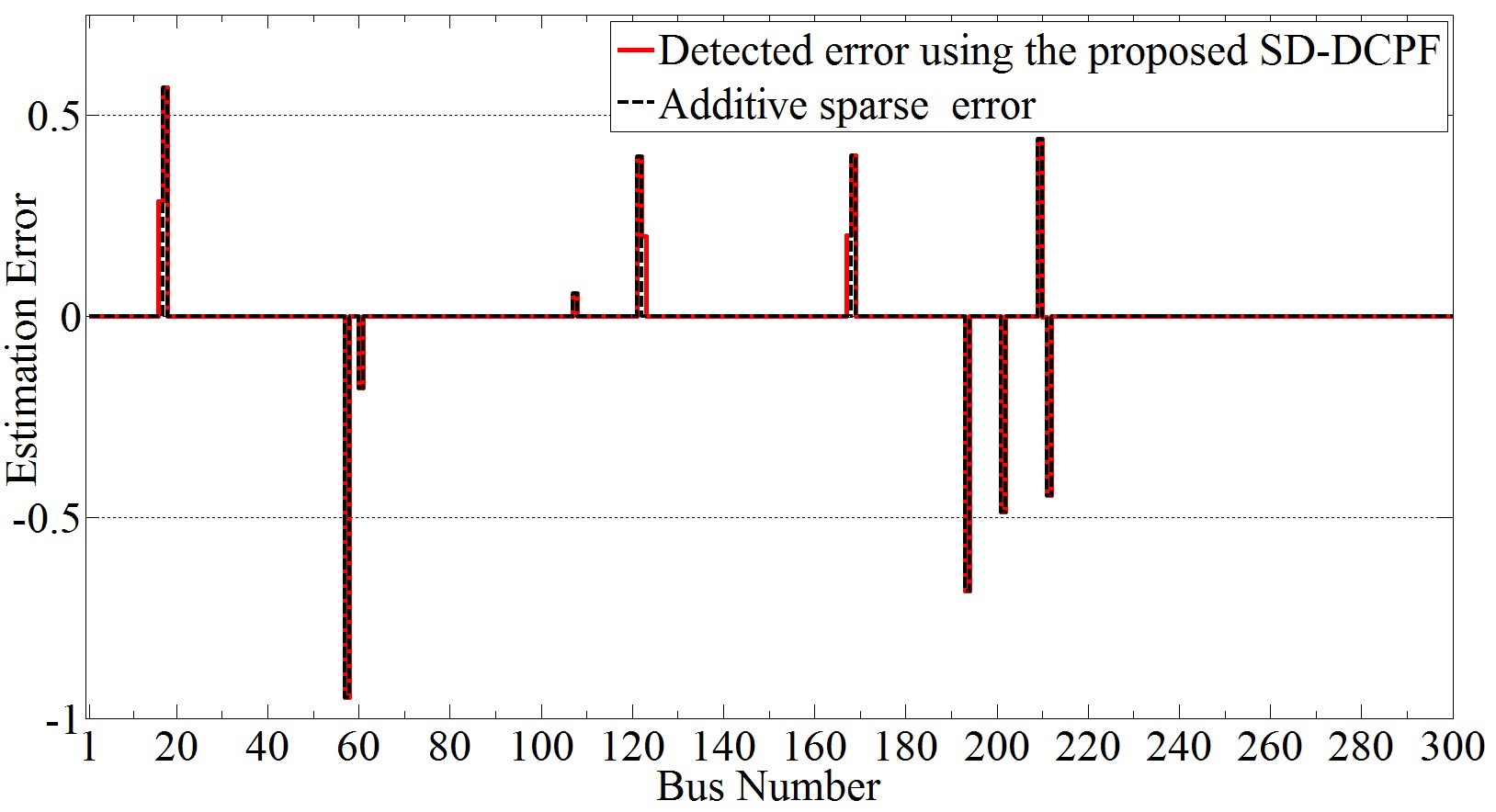}
       \vspace{-.1in}
       \caption{Detected estimation error, \textit{Scenario {IV}}, $\alpha=0.03$.}
       \label{fig:one}
   \end{figure}

\vspace{1cm}
   \noindent  \textit{Scenario V: Comparison between the proposed method and the least-square estimator (LSE):\\}
In this scenario, we assume that the measurements are contaminated with Additive White Gaussian Noise plus highly-sparse injected noise with large magnitude. The sparse noise may be caused by intentional data injection attacks. The result for this scenario, shown in figures 6 and 7, demonstrates the effectiveness of our proposed approach in dealing with noise vectors with elements having large magnitudes. LSE projects the vector $\epsilon$ onto the columns subspace of the matrix \textbf{B}. Thus, the projection is non-negligible when the noise measurement vector has a large norm. In contrast, the $l_1$ norm minimization algorithm is a decomposition algorithm and its performance does not depend on the magnitude of each component.

   \begin{figure}[h!]
   	\centering
       \includegraphics[width=0.5\textwidth]{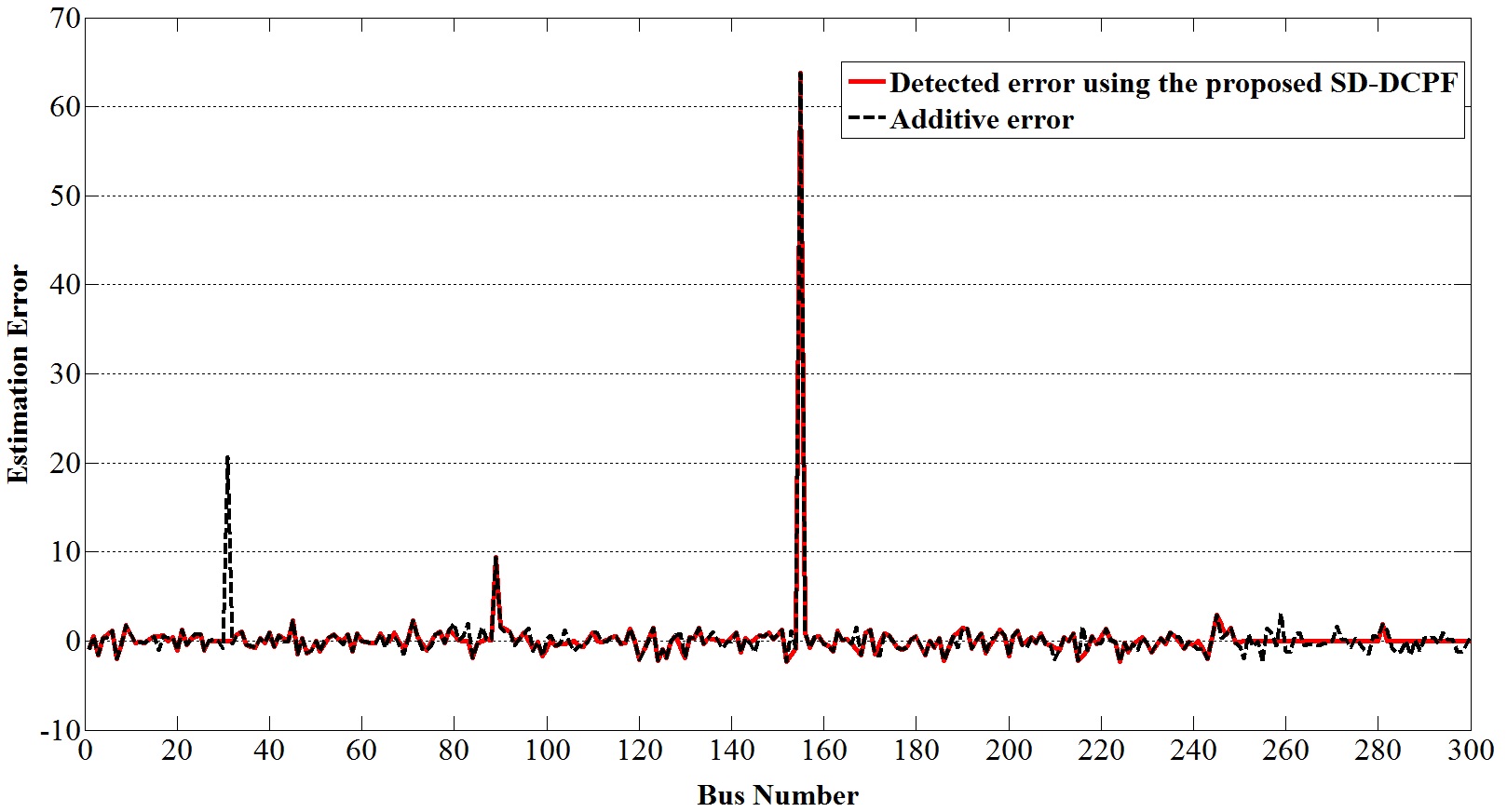}
       \vspace{-.1in}
       \caption{\textit{Scenario {V}}, White Gaussian Noise + Sparse and large magnitude additive noise,  $\alpha_{sparse}=0.02$, utilizing SD-DCPF.}
       \label{fig:one}
   \end{figure}

      \begin{figure}[h!]
      	\centering
          \includegraphics[width=0.5\textwidth]{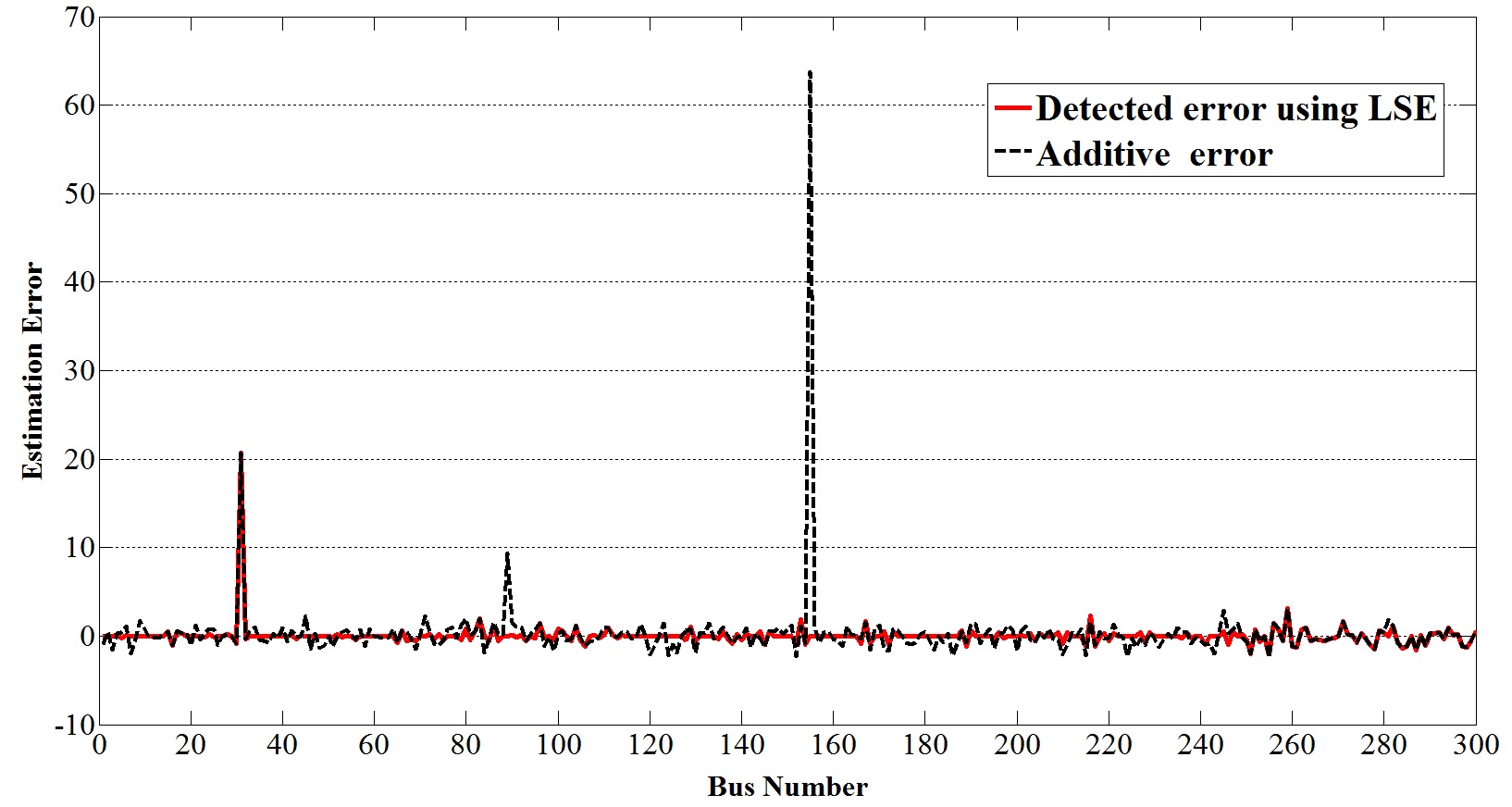}
          \vspace{-.1in}
          \caption{\textit{Scenario {V}}, White Gaussian Noise + Sparse and large magnitude additive noise,  $\alpha_{sparse}=0.02$, utilizing LSE.}
          \label{fig:one}
      \end{figure}
We investigated the effect of $\alpha$ on noise detection for the second test system. It follows the same trend observed in the 118-bus network, i.e. the performance of the proposed SD-DCPF degrades by increasing $\alpha$. Fig. 4 shows the noise detection for scenario IV.

\noindent\textit{Scenario VI: Effect of the magnitude of the noise vector elements on error detection:\\}
We use sparse error elements with values in the range $[-100,100]$ and $\alpha=8\%$. The results validate the robustness of the proposed approach to the magnitude of the noise elements. In this scenario, the error detection rate is $79.16 \%$, i.e. our method detected 19 out of 24 non-zero elements of the noise vector.
            \begin{figure}[h!]
            	\centering
                \includegraphics[width=0.4\textwidth]{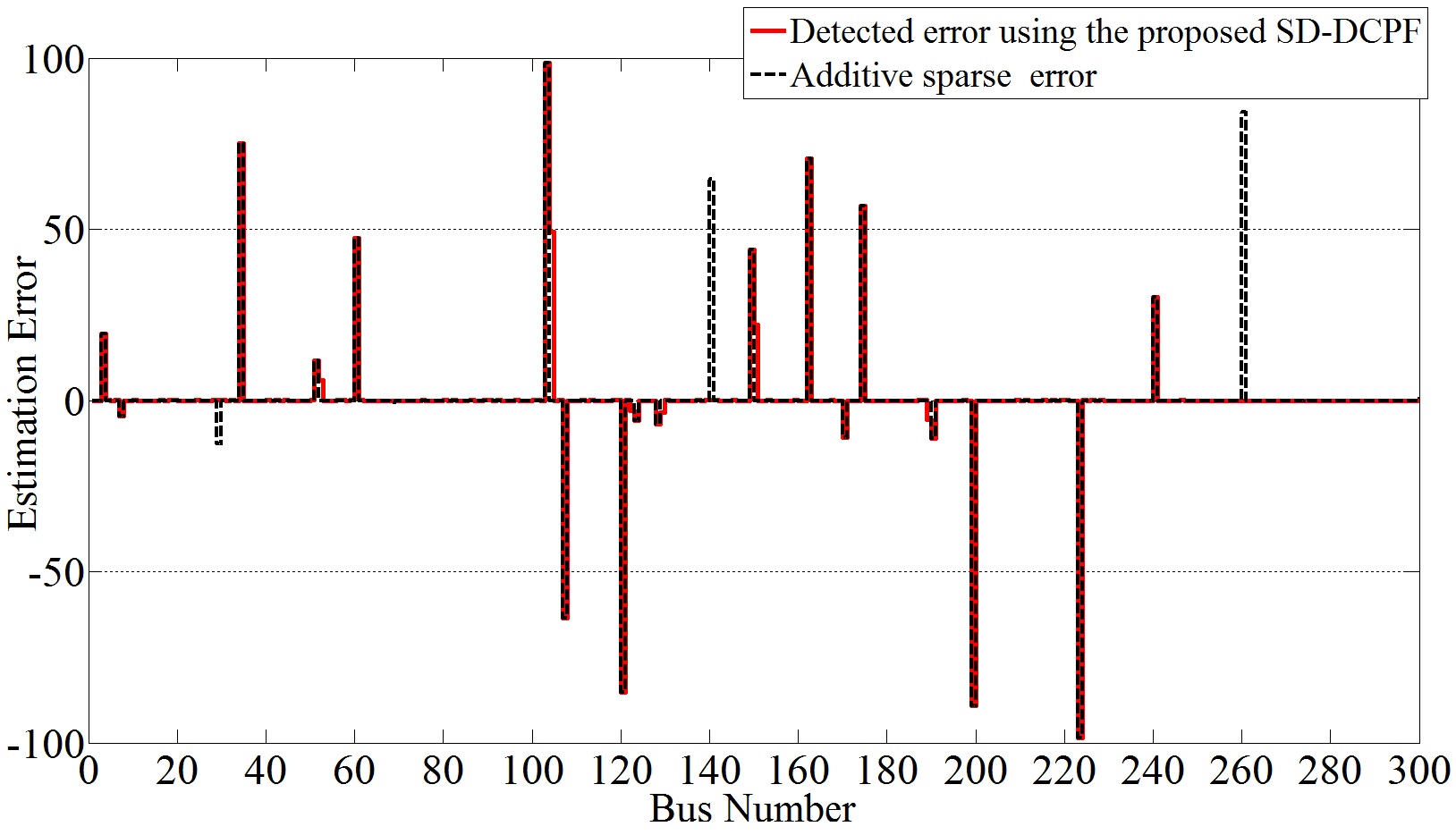}
                \vspace{-.1in}
                \caption{\textit{Scenario {VI}},   $\alpha=0.08$, noise magnitude $ \in [-100,100]$.}
                \label{fig:one}
            \end{figure}


 In order to investigate the performance of the proposed method on different networks, we defined the detection rate as the percentage of detected noisy elements to the total number of injected noisy data elements. Figure 9 shows the results for the analyzed test cases.
 \begin{figure}[h!]
      	\centering
          \includegraphics[width=0.5\textwidth] {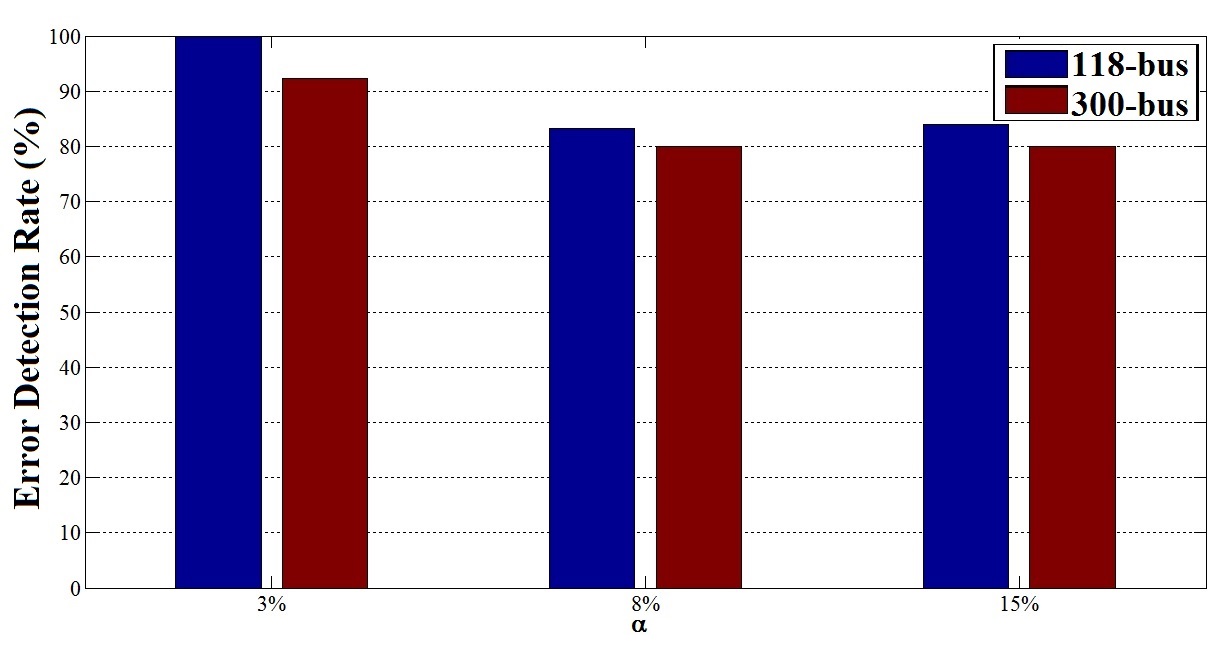} 
          \vspace{-.1in}
          \caption{Error detection rate of SD-DCPF for 118-bus and 300-bus test cases.}
          \label{fig:last}
      \end{figure}
    \section{Conclusion}
An efficient algorithm for identifying the measurement errors in the DC power flow state estimation problem was proposed. In the proposed algorithm, the sparsity of the dominant elements of the error vector and the singularity of the impedance matrix are exploited to make the estimator robust against sparse error elements with arbitrarily large magnitude.
We evaluated the performance of the proposed error detection approach based on different scenarios considering IEEE 118-bus and IEEE 300-bus test systems. 

The detection rate was shown to improve with higher sparsity, i.e., with noise vectors having a smaller support.
Exploiting the grid topology contributes to the accuracy of the proposed method.
To ensure that the proposed decomposition method yields the exact sparse vector, the column subspace of the matrix $\textbf{B}$ should not be aligned with the standard basis.

\bibliographystyle{IEEEtran}
\bibliography{PES_bib_v1}

\end{document}